\documentclass[12pt]{article}

\usepackage{amssymb}
\usepackage{amsfonts}
\usepackage{amsbsy}
\usepackage{amsmath}
\usepackage{amsthm}
\usepackage{graphicx}
\usepackage{epstopdf}
\usepackage[vcentermath]{youngtab}
\usepackage{multirow}
\usepackage{latexsym}
\usepackage{array}
\usepackage{epsfig}

\def\Z{\mathbb{Z}}

\def\n3a{t}

\newcommand{\be}{\begin{equation}}
\newcommand{\ee}{\end{equation}}

\newcommand {\beq}{\begin{equation}}
\newcommand {\eeq}{\end{equation}}
\newcommand {\beqa}{\begin{eqnarray}}
\newcommand {\eeqa}{\end{eqnarray}}
\newcommand {\nn}{\nonumber \\ }

\begin{document}

\begin{titlepage}

\begin{flushright}
APCTP Pre2012-010
\end{flushright}

{\begin{center} {\Large\bf  
Higgsing intersecting brane models on $T^4/\Z_N$}
\end{center}
}

\begin{center}
{\large
Satoshi Nagaoka}\\
Asia Pacific Center for Theoretical Physics\\
Pohang, Gyeongbuk 790-784,
Korea\\
{\tt nagaoka} {\rm at} {\tt apctp.org}
\end{center}

\vspace*{1cm}

\abstract{
We analyze intersecting brane models on toroidal orbifolds
$T^4/\Z_N(N=3,4,6)$. 
Focusing on several Higgsing processes of intersecting brane models,
we have some insights into the matter spectrum on intersecting branes
away from orientifold planes.
The matter fields in the adjoint representations of the gauge group
on branes can arise when brane intersects the orientifold images.
Gauge group and matter spectrum are confirmed using 
Chan-Paton methods with the Higgsing processes.
}

\end{titlepage}

\tableofcontents


\section{Introduction}

There is a huge number of vacua in string theory. 
Orientifolds of type II string theory provide some consistent string theory
vacua. 
Six dimensional supersymmetric gauge theories coupled to gravity 
are constructed by compactifying orientifolds of type II string
theory on a K3 surface.
Intersecting brane models in six dimensions give us 
a tool to describe semi-realistic world since many features
of the models in six dimensions are also relevant in four dimensions.
Initial study of intersecting brane models and their T-dual
magnetized brane models is seen in \cite{Bachas,BGKL,AADS,BKL} and
progress of the construction of vacua with Standard Model gauge group 
and matter content from intersecting branes is seen in 
\cite{CSU1,CSU2,Uranga,Honecker1,HO,Honecker2,IBM-review,GBHLW,BKLS,GH}.

In this paper, we consider 6D intersecting brane models on toroidal 
orbifold $T^4/\Z_N$
where $N=3,4,6$. Anomaly cancellation conditions give us 
a perspective of string landscape in six dimensions
\cite{KT-K3,universality,bound,tensors,0}.
$T^4/\Z_N$ orbifolds are obtained by taking the orbifold limit of K3.
$T^4/\Z_N$ spaces
are locally flat and the analysis is simplified.
In the earlier work \cite{bgk-6D, bbkl,Pradisi,Uranga2,BGK-2,FHS}, 
the gauge group and matter spectrum of intersecting 
brane models on these spaces are studied.
Here we focus on the several Higgsing processes of intersecting brane
models on $T^4/\Z_N$.
Gauge group and the matter spectrum on intersecting branes away
from orientifold planes are obtained in this analysis.
Spectrum of intersecting brane models is not completely
determined by counting the intersection between branes and their
orientifold images
because of the ambiguities \cite{Nagaoka}.
In particular, the contributions to the matter representations
from 
the intersections between the brane and the 
orientifold image away from the orientifold plane
are either a symmetric plus an antisymmetric 
representation or an adjoint representation.
My claim is that in order to determine this ambiguity, 
we need to investigate the Higgsing process.
We have some insights into intersecting brane models on $T^4/\Z_N$.

In Section 2, we review intersecting brane models on
$T^4/\Z_3,\Z_4,\Z_6$.
In Section 3, we analyze Higgsing processes and obtain
gauge group and matter spectrum of intersecting brane models away from 
orientifold planes.
We derive the spectrum using more familiar
Chan-Paton methods.
Section 4 is devoted to the conclusions and discussions.

\section{Intersecting brane models on toroidal orbifolds $T^4/\Z_N$}

Intersecting brane models are constructed by putting branes along 
the cycles in the string compactification manifolds.
Branes are intersecting with each other in the manifolds.
Gauge groups in the space-time theory are produced by  
world-volume gauge fields of D-branes. Matter fields are 
coming from open strings at the intersections of branes.
The low energy theory of intersecting brane models 
can be calculated by choosing 
compactification space such as toroidal orbifolds.
Intersecting D7-brane models with orientifold 7-planes in six dimensions
can be obtained by compactifying type IIB string theory on the K3 surface.
K3 becomes locally flat in the orbifold limit $T^4/\Z_N (N=2,3,4,6)$.

\subsection{Intersecting brane models on $T^4/\Z_3$}

We review intersecting brane models on $T^4/\Z_3$.
We define $T^4=T^2 \times T^2$ through the complex coordinates
$Z_1=X_6+iX_7$, $Z_2=X_8+iX_9$.
There are two possible root lattices which we denote type {\bf A}
and type {\bf B}.
The identifications of the coordinates are
\begin{align}
Z_i \sim Z_i +1 \sim Z_i +e^{\pi i/3} \ ,
\end{align}
for type {\bf A} lattice and
\begin{align}
Z_i \sim Z_i +e^{\pi i /6} \sim Z_i +e^{-\pi i/6} \ ,
\end{align}
for type {\bf B} lattice.
Two $T^2$ take either type {\bf A} or type {\bf B} lattice.
Thus, we have three choices of lattice {\bf AA},{\bf AB},{\bf BB}
which labels the compactification manifolds.
The generators of the $\Z_3$ orbifold group act by
\begin{align}
&\rho : Z_1 \to \exp (2\pi i /3) Z_1 \ , \nn
&\rho : Z_2 \to  \exp {(-2 \pi i /3)} Z_2 \ .
\end{align}
We need orientifold 7-planes in order to include D-branes in a
supersymmetric fashion. 
These are defined by imposing $Z_2 $ symmetry $\Omega \sigma$ on
strings where $\Omega$ is an operator reversing orientation on the
string worldsheet and $\sigma$ is an isometry of the space-time
\begin{align}
\sigma : Z_i \to \bar{Z}_i \ .
\end{align}
This discrete symmetry is realized by putting orientifold 7-planes along $X_6$ and $X_8$.
The orbifold condition produces another discrete symmetries
$\Omega \sigma \rho$ and $\Omega \sigma \rho^2$.

Next, we put D7-branes on the toroidal orbifold.
Tadpole cancellation condition reads
\begin{align}
N=8 \ ,
\end{align}
where N is the number of branes in a stack. 
In this case, the tadpole condition for the $T^4/\Z_3$ 
is a single condition.

There are two distinct possibilities how to put branes in the 
toroidal orientifolds $T^4/\Z_3$:

(i) Branes wrapped on orientifold cycles

(ii) Branes away from orientifold cycles

For the type {\bf AA} lattice, gauge group and matter spectrum arising 
from type (i) branes are
\begin{align} \label{Z3AA}
G=SO(8), \quad {\rm matter}=2 \times A({\bf 28}) \ 
\end{align}
where A is a number of matter fields in the antisymmetric representations.
I will show the derivation of the spectrum using Chan-Paton methods
in Section 3.3.
Gauge group and matter spectrum for the type {\bf AB} lattice 
arising from type (i) branes are
\begin{align}\label{Z3AB}
G=SO(8), \quad {\rm matter}=4 \times A({\bf 28}) \ .
\end{align}
For the type {\bf BB} lattice, gauge group and matter spectrum are
\begin{align}\label{Z3BB}
G=SO(8), \quad {\rm matter}=10 \times A({\bf 28}) \ .
\end{align}
Closed string massless spectrum of each type of lattice is calculated as
\begin{align}
&{\rm 1 SUGRA }+ 8 n_T + 13 n_h \ \   {\rm for}\  {\bf AA}\  {\rm lattice}
 \ , \\
&{\rm 1 SUGRA }+ 6 n_T + 15 n_h \ \   {\rm for}\  {\bf AB}\  {\rm lattice}
 \ , \\
&{\rm 1 SUGRA }+ 21 n_h \ \   {\rm for}\  {\bf BB}\  {\rm lattice} \ ,
\end{align}
where SUGRA, $n_T,n_h$ are numbers of supergravity multiplets,
tensor multiplets and scalar hypermultiplets in the models.
All these models satisfy the anomaly cancellation conditions \cite{gsw}.

\subsection{Intersecting brane models on $T^4/\Z_4$ and $T^4/\Z_6$}

In the case of $T^4/\Z_4$, the identifications of the coordinates are
\begin{align}
Z_i \sim Z_i+1 \sim Z_i +i
\end{align}
for type {\bf A} lattice and
\begin{align}
Z_i \sim Z_i + e^{\pi i/4} \sim Z_i +e^{-\pi i /4}
\end{align}
for type {\bf B} lattice.
Only type {\bf AB} lattice is well-defined model in this case.
The generators of $\Z_4$ orbifold group act by
\begin{align}
&\rho : Z_1 \to \exp (\pi i /2) Z_1 \ , \nn
&\rho : Z_2 \to  \exp {(- \pi i/2 )} Z_2 \ .
\end{align}
The orbifold condition implies that the discrete $\Z_2$ 
produced by $1,\Omega \sigma \rho,\Omega \sigma
\rho^2,\Omega\sigma\rho^3$ are symmetries.
This symmetry is realized by putting
orientifold 7-planes \{a,b,c,d\} as in the figure 1 where
a is stretched along $X_6$ and $X_8$ directions and c is stretched 
along $X_7$ and $X_9$ directions.
b and d are rotated $\pi/4$ from a and c.
In this case, we have two sets of orientifold planes which are not related by
$\Z_4$. One set is orientifold planes $\{a,c\}$ and 
the other set is orientifold planes $\{b,d\}$.

\begin{figure}[t]
\begin{center}
\includegraphics[width=6cm]{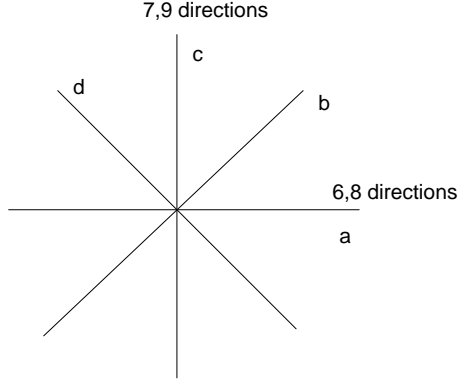}
\caption{Orientifold planes \{$a,b,c,d$\} are located on $Z_i(i=1,2)$.}
\label{fig1}
\end{center}
\end{figure}

In order to cancel Ramond-Ramond (RR) charges of orientifold planes,
we need to introduce two sets of D7-branes.
Tadpole cancellation condition gives 
\begin{align}
N_1=8, \quad N_2=8
\end{align}
where $N_1$ and $N_2$ are numbers of branes to cancel the RR charge of 
the sets $\{a,c\}$
and $\{b,d\}$.
Gauge group and matter spectrum arising from type (i) branes are
\begin{align}
G=U(8) \times U(8), \quad {\rm matter}=4 \times (A,1) + 4 \times 
(1,A)+2 \times (F,F) \ .
\end{align}

In the case of $T^4/\Z_6$, 
the identifications of the coordinates are
\begin{align}
Z_i \sim Z_i +1 \sim Z_i +e^{\pi i/3} \ ,
\end{align}
for type {\bf A} lattice and
\begin{align}
Z_i \sim Z_i +e^{\pi i /6} \sim Z_i +e^{-\pi i/6} \ ,
\end{align}
for type {\bf B} lattice.
The generators of the $\Z_6$ orbifold group act by
\begin{align}
&\rho : Z_1 \to \exp (\pi i /3) Z_1 \ , \nn
&\rho : Z_2 \to  \exp {(- \pi i /3)} Z_2 \ .
\end{align}
Six orientifold 7-planes are located at
\{a,b,c,d,e,f\} where
a is stretched along $X_6$ and $X_8$ directions and 
b is rotated $\pi/6$ from a and 
c is rotated $\pi/6$ from b, and so on.
In this case, we have two sets of orientifold planes which are not related by
$\Z_6$. One set is orientifold planes $\{a,c,e\}$ and 
the other set is orientifold planes $\{b,d,f\}$.

We consider {\bf AB} lattice. 
Tadpole cancellation condition gives
\begin{align}
N_1=4, \quad N_2=4
\end{align}
where $N_1,N_2$ are the numbers of branes in each set.
Gauge group and matter spectrum are
\begin{align}
G=U(4) \times U(4)\ , \quad  {\rm matter}= 6 \times (A,1) + 6 \times (1,A)
+({\rm adjoint},1)+(1,{\rm adjoint})+6 \times (F,F) \ 
\end{align}
where $F$ is matter fields in the fundamental representations.

The story so far has been a summary of intersecting brane models on
$T^4/\Z_3,\Z_4,\Z_6$ \cite{bgk-6D, bbkl,Pradisi,Uranga2,BGK-2,FHS}.

\section{Higgsing intersecting brane models on $T^4/\Z_N$}

\subsection{Higgsing intersecting brane models}

Type (ii) branes appear when we move type (i) branes away from orientifold
planes. This is caused by Higgsing,
turning on a VEV for some scalar fields.
Gauge group and matter spectrum of type (ii) branes for {\bf AA} lattice
are 
\begin{align}
G=SU(4), \quad {\rm matter}=2 \times A({\bf 6}) + 2 \times {\rm
 adjoint}({\bf 15})\ .
\end{align}
Under the decomposition $SU(N) \subset SO(2N)$, the matter representation
in $SO(2N)$ group becomes \cite{Slansky}
\begin{align}
A_{SO(2N)} \rightarrow 2A_{SU(N)} \oplus D_{SU(N)} \ .
\end{align}
In this case,
Higgsing is performed by putting on VEV for two $A_{SU(N)}$,
so totally, we obtain the massless matter fields
in two antisymmetric representations and two adjoint representations.
The gauge group and matter content can also be obtained by 
the analysis using Chan-Paton methods, which I show in Section 3.3.

In the same way, we can obtain spectrum for {\bf AB} and {\bf BB} lattice.
For {\bf AB} lattice, 
gauge group and matter spectrum of type (ii) branes are 
\begin{align}
G=SU(4), \quad {\rm matter}=6 \times A({\bf 6}) + 4 \times {\rm adjoint}
({\bf 15})\ .
\end{align}
For {\bf BB} lattice, 
gauge group and matter spectrum of type (ii) branes are 
\begin{align}
G=SU(4), \quad {\rm matter}=18 \times A({\bf 6}) + 10 \times {\rm
 adjoint}
({\bf 15})\ .
\end{align}
The anomaly cancellation conditions are satisfied for all of 
these models.

Gauge group and matter spectrum from
branes away from orientifold planes (type (ii)) on $T^4/\Z_4$ are
\begin{align}
G=Sp(4) \times Sp(4), \quad {\rm matter}=2 \times (A,1) +2 \times
(1,A) +2 \times (F,F) \ .
\end{align}

Gauge group and matter spectrum from
branes away from orientifold planes on $T^4/\Z_6$ are
\begin{align}
G=Sp(2) \times Sp(2), \quad {\rm matter}=6 \times (A,1) +6 \times
(1,A) + (S,1)+(1,S)+
6 \times (F,F) \ ,
\end{align}
where $S$ is matter fields in the symmetric representations.

\subsection{Chan-Paton analysis}

Gauge symmetry and matter spectrum of intersecting brane models are
determined by the analysis using Chan-Paton factor.
We introduce a Chan-Paton matrix $\lambda_{ij}$ and general
open string states $|\psi,ij \rangle$ where $\psi$ is the state of the
world-sheet fields.
By the action of $\Omega$, the open strings are translated into 
\begin{equation} \Omega \lambda_{ij} |\psi , ij \rangle 
\to \lambda_{ij} (\gamma_{\Omega})_{ii'}
|\Omega \cdot \psi, j'i' \rangle (\gamma_\Omega^{-1})_{j'j}
\end{equation}
where $\gamma_\Omega$ is some matrix.
$\Omega^2$ acts on the string world-sheet as 
\begin{equation}
\Omega^2 : |\psi , ij \rangle \to
(\gamma_{\Omega }(\gamma_{\Omega }^T)^{-1})_{ii'}
|\psi,i'j' \rangle (\gamma_{\Omega }^T
\gamma_{\Omega }^{-1})_{j'j} \ .
\end{equation}
Since this operator
acts trivially
 on the world-sheet fields,
we obtain
\begin{equation}
\gamma_{\Omega }^T=\pm \gamma_{\Omega } \ .
\end{equation}

We start from type (i) brane on $AA$ lattice of $T^4/\Z_3$.
We can separate the Chan-Paton matrix into $3\times 3$ blocks
which labels open strings on branes and 
their orientifold images $(1,\Omega \sigma)$ and 
orbifold images of the pair $(\rho,\rho\Omega\sigma)$ and $(\rho^2,\rho^2
\Omega\sigma$). 

Since gauge fields coming from the orbifold image are equivalent to 
gauge fields from branes,
the Chan-Paton matrix for the gauge fields can be separated into
\begin{equation}  \label{AA3block}
\lambda_{{\rm whole}}= \left( \begin{array}{ccc} \lambda&&  \\ &\lambda&
\\ &&\lambda
 \end{array}
\right) \ 
\end{equation}
where the off-diagonal blocks are not relevant for the massless
excitations.
Since branes are located on top of orientifold planes,
$\lambda$
satisfies
\begin{align} \label{CPgauge1}
\lambda =- \gamma_\Omega \lambda^T \gamma_\Omega
\end{align}
where $\gamma_\Omega$ is taken to be
\begin{align}
\gamma_\Omega=I \ .
\end{align}
Thus, we obtain SO(8) gauge group from $(\ref{CPgauge1})$.

In this case, the intersections between branes and orientifold images are
located on top of orientifold planes. They give a single matter fields
in the antisymmetric representations.
The Chan-Paton matrix for 
each diagonal block of matter fields satisfies
\begin{align}
\lambda=-\lambda^T \ ,
\end{align}
which gives a single antisymmetric representation.
The total Chan-Paton matrix is obtained as
\begin{align}
\lambda_{{\rm whole}}= \left( \begin{array}{ccc} A_1&A_2&A_2  \\ A_2&A_1&A_2
\\ A_2&A_2&A_1
 \end{array}
\right) \ .
\end{align}
Thus, we have obtained SO(8) gauge group with 2 matter fields in the 
antisymmetric representations.

For the {\bf AB} lattice, the number of intersection between branes and 
orientifold images becomes three times of the {\bf AB} lattice.
Thus, we obtain $1+3=4$ antisymmetric representations.
 For the {\bf BB}
lattice, the number of intersections is nine. Thus,
we obtain $1+9=10$ antisymmetric representations.

Next, we consider type (ii) branes on {\bf AA} lattice of $T^4/\Z_3$.
Since all images are not located on top of each other,
we separate Chan-Paton matrix into $6\times 6$ blocks
which label open strings on 
$(1, \Omega \sigma, \rho,\rho\Omega\sigma ,\rho^2,\rho^2
\Omega\sigma)$. 
Since all images represent the same strings, 
the solution of Chan-Paton matrix for gauge fields is 
\begin{align}
\lambda_{{\rm whole}}= \left( \begin{array}{cccccc}D&&&&&   \\ &D&&&&
\\ &&D&&& \\&&&D&& \\&&&&D& \\&&&&&D
 \end{array}
\right) \ 
\end{align}
where D is a $4\times 4$ hermitian block representing the SU(4) gauge
 group.
Chan-Paton matrix for matter fields is obtained in the same way as
\begin{align} \label{AA6block}
\lambda_{{\rm whole}}= \left( \begin{array}{cccccc}D_1&&&A_1&&D_2   
\\ &D_1&A_2&&D_2&
\\ &A_2&D_1&&&A_1 \\A_1&&&D_1&A_2& 
\\&D_2&&A_2&D_1& \\D_2&&A_1&&&D_1
 \end{array}
\right) \ .
\end{align}
An adjoint representation $D_2$ comes from the intersection of three
images where two of them are exchanged under the action 
$\Omega$, but one is invariant under the action $\Omega$ \cite{bgk-6D}.
The latter one gives a matter field either in 
the adjoint representation or
in
the single symmetric representation plus antisymmetric representation. 
This ambiguity is determined completely
by considering the Higgsing process \cite{Nagaoka}.
In this case, (\ref{AA6block}) is consistent with the decomposition
\begin{align}
A_{SO(8)} \to 2A_{SU(4)} \oplus D_{SU(4)} \ .
\end{align}

Thus,
we have obtained the matter spectrum
\begin{align}
{\rm matter}= 2 \times A + 2 \times {\rm adjoint} \ .
\end{align}

Note that a block in the (\ref{AA3block}) is separated into $2\times
2$
blocks in the (\ref{AA6block}). This means that an intersection 
in type (i) brane is 
decomposed into a group of four intersections in type (ii) brane. 
The group of four intersections gives 2 matter fields in the
antisymmetric representations and an adjoint matter representation.

For the {\bf AB} lattice, there are three intersections between branes and
their orientifold images. Since the matter spectrum coming from 
each intersection is 
2 antisymmetric representations and 1 adjoint representation,
this intersecting brane model has SU(4) gauge group with 6 matter fields
in the antisymmetric representations and 4 matter fields in the
adjoint representations.

For the {\bf BB} lattice of $T^4/\Z_3$, the number of intersections
between
branes and their orientifold images is nine. 
Thus, this model has SU(4) gauge group 
with 18 matter fields in the antisymmetric representations and
10 matter fields in the adjoint representations.

In the case of $T^4/\Z_4$, there are two sets of branes.
Chan-Paton matrix is separated into $2\times
2$ blocks 
\begin{align}
\lambda_{{\rm whole}}= \left( \begin{array}{cc} \lambda_1&0  \\ 0&\lambda_2
 \end{array}
\right) \ 
\end{align}
where each $\lambda_i(i=1,2)$ is Chan-Paton matrix of each set of branes.
$\lambda_1$ labels strings on branes stretched along 6,8 directions (a
in the figure 1) and $\lambda_2$ labels strings on branes rotated
$\pi/4$ from a (b in the figure 2).
In order to solve $\lambda_i$, we separate
\begin{align}
\lambda_{i}= \left( \begin{array}{cc} \lambda_i^{11}&\lambda_i^{12}
  \\ \lambda_i^{21}&\lambda_i^{22}
 \end{array}
\right) \ 
\end{align}
where
$\lambda_i^{11}$  labels strings on branes 
$(1,\Omega\sigma,\rho^2,\rho^2 \Omega\sigma)$ and $\lambda_i^{22}$ labels
strings on branes 
 $(\rho,\Omega\sigma\rho,\rho^3,\Omega\sigma\rho^3)$.
Off-diagonal modes have no
massless excitations, that is, $\lambda_i^{12}=\lambda_i^{21}=0$.
In this case, branes, their orientifold images $\Omega\sigma$,
their orbifold images $\rho^2$ and $\rho^2\Omega\sigma$ are
on top of each other.
The Chan-Paton matrix of gauge fields $\lambda_1^{11}$ satisfies 
\cite{GP,Nagaoka}
\begin{align} 
\lambda_1^{11}=+\gamma_{\rho^2} \lambda_1^{11} \gamma_{\rho^2}^{-1}
\ , \quad 
\lambda_1^{11}=-\gamma_{\Omega\sigma} (\lambda_1^{11})^T 
\gamma_{\Omega\sigma} \ . \label{cpcond1}
\end{align}
By a unitary transformation, we can take 
\begin{align} 
\gamma_{\Omega \sigma}=I, \quad \gamma_{\rho^2}
 =\gamma_{\Omega\sigma\rho^2}= 
\left( \begin{array}{cc} 0&iI/2  \\ -iI/2&0
 \end{array}
\right) \ .
\end{align}
Chan-Paton matrix of gauge fields $\lambda_2^{22}$ satisfies
\begin{align}
\lambda_2^{22}=+\gamma_{\rho^3} \lambda_2^{22} \gamma_{\rho^3}^{-1}
\ , \quad 
\lambda_2^{22}=-\gamma_{\Omega\sigma\rho} (\lambda_2^{22})^T 
\gamma_{\Omega\sigma\rho}
\ . \label{cpcond2}
\end{align}
By a unitary transformation, we can take
\begin{align}
\gamma_{\Omega \sigma\rho}=I, \quad \gamma_{\rho^2}
 =\gamma_{\Omega\sigma\rho^3}= 
\left( \begin{array}{cc} 0&iI/2  \\ -iI/2&0
 \end{array}
\right) \ .
\end{align}

By solving the condition (\ref{cpcond1}) and (\ref{cpcond2}), 
we obtain the Chan-Paton matrices of
gauge fields,
\begin{align}
\lambda_1^{11}= \left( \begin{array}{cc} A_1&S_1  \\ -S_1&A_1
 \end{array}
\right) \ , \quad
\lambda_2^{22}= \left( \begin{array}{cc} A_2&S_2  \\ -S_2&A_2
 \end{array}
\right) \ .
\end{align}
Since $\lambda_1^{22}(\lambda_2^{11})$ is an orbifold image of 
$\lambda_1^{11}(\lambda_2^{22})$, they do not produce further gauge group.
Thus, we have obtained $U(8)\times U(8)$ gauge group.

The Chan-Paton matrices of matter fields satisfy \cite{GP,Nagaoka}
\begin{align}\nonumber
&\lambda_1^{11}=-\gamma_{\rho^2} \lambda_1^{11} \gamma_{\rho^2}^{-1}
\ , \quad 
\lambda_1^{11}=-\gamma_{\Omega\sigma} (\lambda_1^{11})^T 
\gamma_{\Omega\sigma} \ , \\
&\lambda_2^{22}=-\gamma_{\rho^3} \lambda_2^{22} \gamma_{\rho^3}^{-1}
\ , \quad 
\lambda_2^{22}=-\gamma_{\Omega\sigma\rho} (\lambda_2^{22})^T 
\gamma_{\Omega\sigma\rho}
 \ .
\label{cpcondm}
\end{align}
By solving the condition (\ref{cpcondm}), 
we obtain the Chan-Paton matrix of
matter fields.
\begin{align}
\lambda_1^{22}= \left( \begin{array}{cc} A_1&A_2  \\ A_2&-A_1
 \end{array}
\right) \ , \quad
\lambda_2^{22}= \left( \begin{array}{cc} A_1&A_2  \\ A_2&-A_1
 \end{array}
\right) \ .
\end{align}
Thus, we have obtained two matter fields in the antisymmetric
representations for each set of branes.
There are two intersections for each $\lambda_i$. 
In addition, there are two intersections between different set of
branes which produce two matter fields in the bifundamental representations.
 Thus, 
we have matter fields
\begin{align}
{\rm matter}=4\times(A,1)+4 \times(1,A)+2\times(F,F) \ .
\end{align}

Gauge group and matter fields of type (ii) brane on $T^4/\Z_4$
are obtained by Higgsing type (i) brane.
Each block of $\lambda_i^{11}, \lambda_i^{22},\lambda_i^{33}$ and 
$\lambda_i^{44}$ labels
open strings on branes
$(1,\rho^2\Omega\sigma)$,
$(\Omega\sigma,\rho^2)$,
$(\rho,\Omega\sigma\rho^3)$ and $(\Omega\sigma\rho,\rho^3)$.
\begin{align}
\lambda_i=
\left( \begin{array}{cccc} \lambda_i^{11}&&&  \\ &\lambda_i^{22}&& 
\\ &&\lambda_i^{33}& \\ &&&\lambda_i^{44}
 \end{array}
\right) \ .
\end{align}
Since branes are located on top of orientifold planes,
we obtain $Sp(4)$ gauge group for type (ii) brane.

The solution of the Chan-Paton matrix for the matter fields is
\begin{align}
\lambda_1=
\left( \begin{array}{cccc} &A&A&A  \\ A&&A&A 
\\ A&A&&A \\ A&A&A&
 \end{array}
\right) \ ,
\end{align}
which gives a matter field in the 
single antisymmetric representation to each set of branes.
Since there are two intersections for each set of branes,
we have the matter spectrum and gauge group
\begin{align}
G=Sp(4) \times Sp(4), \quad
{\rm matter} = 2 \times (A,1) +2 \times (1,A) + 2 \times (F,F) \ .
\end{align}

Finally, we consider $T^4/\Z_6$. 
In this case, two sets of branes can exist. Chan-Paton matrix is
separated into $2\times 2$ blocks
\begin{align}
\lambda_{{\rm whole}}= \left( \begin{array}{cc} \lambda_1&0  \\ 0&\lambda_2
 \end{array}
\right) \ 
\end{align}
where each $\lambda_i(i=1,2)$ is Chan-Paton matrix of each set of branes.
$\lambda_1$ labels strings on branes stretched along 6,8 directions and
 $\lambda_2$ labels strings on branes rotated $\pi/6$ from a.
In order to solve $\lambda_i$, we separate Chan-Paton matrix into
$3\times 3$ blocks
\begin{align}
\lambda_{i}= \left( \begin{array}{ccc} \lambda_i^{11}
&\lambda_i^{12}&\lambda_i^{13}
  \\ \lambda_i^{21}&\lambda_i^{22}&\lambda_i^{23}
\\\lambda_i^{31}&\lambda_i^{32}&\lambda_i^{33}
 \end{array}
\right) \ 
\end{align}
where
$\lambda_i^{11}$  labels strings on branes 
$(1,\Omega\sigma,\rho^3,\rho^3 \Omega\sigma)$ and $\lambda_i^{22}$ labels
strings on branes 
 $(\rho,\Omega\sigma\rho,\rho^4,\Omega\sigma\rho^4)$
and $\lambda_i^{33}$ labels strings on branes 
$(\rho^2,\Omega\sigma\rho^2,\rho^5,\Omega\sigma\rho^5)$.
Off-diagonal modes have no
massless excitations.
The Chan-Paton matrix of gauge fields $\lambda_1$ satisfies \cite{GP,Nagaoka}
\begin{align} 
\lambda_1^{11}=+\gamma_{\rho^3} \lambda_1^{11} \gamma_{\rho^3}^{-1}
\ , \quad 
\lambda_1^{11}=-\gamma_{\Omega\sigma} (\lambda_1^{11})^T 
\gamma_{\Omega\sigma} \ , \label{cpcond3}
\end{align}
By a unitary transformation, we can take 
\begin{align} 
\gamma_{\Omega \sigma}=I, \quad \gamma_{\rho^3}
 =\gamma_{\Omega\sigma\rho^3}= 
\left( \begin{array}{cc} 0&iI/2  \\ -iI/2&0
 \end{array}
\right) \ .
\end{align}
Chan-Paton matrix of gauge fields $\lambda_2$ satisfies
\begin{align}
\lambda_2^{22}=+\gamma_{\rho^4} \lambda_2^{22} \gamma_{\rho^4}^{-1}
\ , \quad 
\lambda_2^{22}=-\gamma_{\Omega\sigma\rho} (\lambda_2^{22})^T 
\gamma_{\Omega\sigma\rho}
\ . \label{cpcond4}
\end{align}
By a unitary transformation, we can take
\begin{align}
\gamma_{\Omega \sigma\rho}=I, \quad \gamma_{\rho^4}
 =\gamma_{\Omega\sigma\rho^4}= 
\left( \begin{array}{cc} 0&iI/2  \\ -iI/2&0
 \end{array}
\right)\ .
\end{align}

By solving the condition (\ref{cpcond3}) and (\ref{cpcond4}), 
we have obtained the Chan-Paton matrices of
gauge fields,
\begin{align}
\lambda_1^{11}= \left( \begin{array}{cc} A_1&S_1  \\ -S_1&A_1
 \end{array}
\right) \ , \quad
\lambda_2^{22}= \left( \begin{array}{cc} A_2&S_2  \\ -S_2&A_2
 \end{array}
\right) \ .
\end{align}
Since $\lambda_1^{22}$ and $\lambda_1^{33}$
($\lambda_2^{11}$ and $\lambda_2^{33}$) are orbifold images of 
$\lambda_1^{11}(\lambda_2^{22})$, they do not produce further gauge group.
Thus, we have obtained $U(4)\times U(4)$ gauge group.

The solution of the Chan-Paton matrix for the matter fields is
\begin{align} 
\lambda_{{\rm whole}}= 
\left( \begin{array}{cccccc}A_1&A_2&&A_3&&D   
\\ A_2&-A_1&A_4&&D&
\\ &A_4&A_1&A_2&&A_5 \\A_3&&A_2&-A_1&A_6& 
\\&D&&A_6&A_1&A_2 \\D&&A_5&&A_2&-A_1
 \end{array}
\right) \ .
\end{align}
An adjoint representation comes from the intersection of three
images of branes where two of them are exchanged under the action 
$\Omega$, but one is invariant under the action $\Omega$ \cite{bgk-6D}.
The latter one gives a matter field in the adjoint representation.
There are six intersections between different set of branes
which produce six matter fields in the bifundamental representations.
Thus, we have obtained the matter fields
\begin{align}
{\rm matter}= 6 \times (A,1) + 6 \times (1,A)
+({\rm adjoint},1)+(1,{\rm adjoint})+6 \times (F,F) \ .
\end{align}

Gauge group and matter fields of type (ii) brane on $T^4/\Z_6$
are obtained by Higgsing type (i) brane.
Each block of $\lambda_1^{ii}(i=1,\cdots,6)$ labels
open strings on branes
$(1,\rho^3\Omega\sigma)$,
$(\Omega\sigma,\rho^3)$,
$(\rho,\Omega\sigma\rho^4)$,$(\Omega\sigma\rho,\rho^4)$
$(\rho^2,\Omega\sigma\rho^5)$,$(\Omega\sigma\rho^2,\rho^5)$.
\begin{align}
\lambda_1=
\left( \begin{array}{cccccc} \lambda_1^{11}&&&&&  \\ &\lambda_1^{22}&&&& 
\\ &&\lambda_1^{33}&&& \\ &&&\lambda_1^{44}&&
\\ &&&&\lambda_1^{55}& \\ &&&&&\lambda_1^{66}
 \end{array}
\right) \ .
\end{align}
Since branes are located on top of orientifold planes,
we obtain $Sp(2)$ gauge group for type (ii) brane.

The solution of the Chan-Paton matrix for the matter fields is
\begin{align}
\lambda_{{\rm whole}}= 
\left( \begin{array}{cccccc}A_1&&&A_3&&S+A_2   
\\ &-A_1&A_4&&S+A_2&
\\ &A_4&A_1&&&A_5 \\A_3&&&-A_1&A_6& 
\\&S+A_2&&A_6&A_1& \\S+A_2&&A_5&&&-A_1
 \end{array}
\right) \ .
\end{align}
The matter fields in the symmetric representations appear by the
decomposition
\begin{align}
A_{U(4)} \rightarrow A_{Sp(2)} \oplus S_{Sp(2)} \ .
\end{align}

Thus, we have the matter spectrum and gauge group
\begin{align}
G=Sp(2) \times Sp(2), \quad
{\rm matter} = 6 \times (A,1) +6 \times (1,A) + (1,S)+(S,1)+6 \times (F,F) \ .
\end{align}

\section {Conclusions and discussions}

We have analyzed the Higgsing processes of intersecting brane models 
on $T^4/\Z_N$. The matter fields in the adjoint representations appear
branes away from orientifold planes (type (ii) brane) in the case of
$T^4/\Z_3$. The matter spectrum and gauge group of branes away from 
orientifold planes have been newly found in the case of $T^4/\Z_4$ and 
$T^4/\Z_6$. Gauge group and matter spectrum have been confirmed using
Chan-Paton methods.

Possible intersecting brane models on $T^4/\Z_N(N=3,4,6)$
are not so much compared to the models on $T^4/\Z_2$ \cite{Nagaoka},
since
we can put branes non-parallel to orientifold cycles on $T^4/\Z_2$.
In the case of $T^4/\Z_3$,
we have one set of orientifold planes which is related by $\Z_3$.
In order to cancel RR-charge of orientifold planes,
 we need to introduce one set of D7-branes, which produces one simple 
gauge group.
On the other hand, we have two sets of orientifold planes
which are not related by $\Z_4$ and $\Z_6$.
Two sets of D-branes give product of the simple gauge group $G=G_1
\times G_2$. Higgsing to type (ii) brane determines the ambiguous 
part of the spectrum. Especially, 
the matter fields in the adjoint representations of the gauge group
on branes can arise when brane intersects the orientifold images.

\end{document}